\documentclass[aps,prl,twocolumn,showpacs,superscriptaddress,amsmath,amssymb]{revtex4-2}

\usepackage{graphicx}
\usepackage{dcolumn}
\usepackage{bm}
\usepackage{amsmath}
\usepackage{amssymb}
\usepackage{graphicx}
\usepackage{indentfirst}
\usepackage{booktabs}
\usepackage{multirow}
\usepackage{colortbl}
\usepackage{epsfig}

\begin{document}

\title{Unusual anisotropic magnetoresistance due to magnetization-dependent spin-orbit interactions}

\author{M. Q. Dong}
\affiliation{State Key Laboratory for Mechanical Behavior of Materials, Center for Spintronics and Quantum System, School of Materials Science and Engineering, Xi’an Jiaotong University, Xi’an, Shaanxi, 710049, China.}
\author{Zhi-Xin Guo}
\email{zxguo08@xjtu.edu.cn}
\affiliation{State Key Laboratory for Mechanical Behavior of Materials, Center for Spintronics and Quantum System, School of Materials Science and Engineering, Xi’an Jiaotong University, Xi’an, Shaanxi, 710049, China.}
\author{X. R. Wang}
\email{phxwan@ust.hk}
\affiliation{Physics Department, The Hong Kong University of Science and Technology, Clear Water Bay, Kowloon, Hong Kong.}
\affiliation{HKUST Shenzhen Research Institute, Shenzhen 518057, China.}

\date{\today}

\begin{abstract}

One of recent surprising discoveries is the unusual anisotropic magnetoresistance (UAMR) that depends on two magnetization components perpendicular to the current differently, in contrast to the conventional anisotropic magnetoresistance (AMR) that predicts no change in resistance when the magnetization varies in the plane perpendicular to the current. Using density functional theory and Boltzmann transport equation calculations for bcc Fe, hcp Co, and bcc FeCo alloys, we show that UAMR can be accounted by the magnetization-dependent spin-orbit interactions (SOI): Magnetization-dependent SOI modifies electron energy bands that, in turn, changes resistance. A phenomenological model reveals the intrinsic connection between SOI and order-parameters. Such a mechanism is confirmed by the strong biaxial stain effect on UAMR. Our findings provide an efficient way of searching and optimizing materials with large UAMR, important in the design of high-performance spintronic devices. 

\end{abstract}

\maketitle


\section{\label{sec1}introduction}

The spin-dependent transport of magnetic materials is an important topic in spintronics \cite{cite1}. The electronic conductivity of a magnetic device usually depends on its magnetization structures, resulting in various types of galvanomagnetic effects. Among these effects, anisotropic magnetoresistance (AMR) of magnetic materials is a well-known phenomenon, which says that the longitudinal electronic resistivity depends on the magnetization direction relative to the electric current
\cite{cite2,cite3,cite4,cite5,cite6,cite7,cite8,cite9,cite10,cite11}.
The AMR is attributed to the relativistic spin-orbit interactions (SOI), which couple the electron orbital motions with their spin angular momentum, and has a universal form in magnetic polycrystals: the change of longitudinal resistivity (or resistance)  $\Delta\rho_{xx}\left(\alpha\right)$ follows 
$\Delta\rho_{xx}\left(\alpha\right)=\Delta\rho_{xx}^0cos^2\alpha$ 
with $\alpha$ being the angle between the magnetization and electric current \cite{cite12}.

Recently, an intrinsic unusual anisotropic magnetoresistance (UAMR) effect has been discovered in FeCo alloy, where the electronic conductivity depends not only on the angle between current and magnetization, but also on the angle between crystalline axis and magnetization \cite{cite13}. Based on symmetry argument \cite{cite14} and tensor analysis \cite{cite15}, Wang revealed the relationship among electronic conductivity/resistivity, order parameters (e.g. magnetization and crystalline axis), and electric current in a phenomenological way \cite{cite16}, which gives rise to an intrinsic UAMR. Nevertheless, the microscopic mechanism for the order-parameter dependent UAMR is yet to be explored.

In this paper, we use the density functional theory (DFT) and Boltzmann transport equations (BTE) to explore the microscopic origin of UAMR by studying several magnetic single crystals including bcc Fe, hcp Co and bcc FeCo alloys. For the electric current along [110] direction (denoted as \textit{x} axis), we consider the magnetization along $\left[\bar{1}10\right]$ (denoted as \textit{y} axis) or along [001] direction (denoted as \textit{z} axis). The longitudinal electronic resistance $\left(\rho_{xx}\right)$ is the same for magnetization along \textit{y} and \textit{z} directions according to the conventional AMR. Whereas they are different according to the UAMR. We show that whether the $\rho_{xx}$ of these crystals varies with the magnetization direction in the \textit{yz}-plane depends on whether the SOI is included. Our calculations demonstrate that UAMR comes from the magnetization-dependent SOI. SOI leads to the energy band splitting near the Fermi level $\left(E_F\right)$ that, in turn, affects electronic transport. This intrinsic mechanism is confirmed by the strong biaxial stain dependence of UAMR.

\section{\label{sec:level2}method}

Our \textit{ab initio} calculations are performed by using the QUANTUM ESPRESSO package based on the projector-augmented wave (PAW) method and a plane-wave basis set \cite{cite17,cite18}. The exchange and correlation terms are described by a generalized gradient approximation (GGA) in the scheme of Perdew-Burke-Ernzerhof (PBE) parametrization, as implemented in the PSLIBRARY \cite{cite19}. Energy convergence criteria of all the calculations is set as $1.0\times10^{-8}$ Ry, and wave-function cutoff of all calculations are 180Ry. In the self-consistent field (scf) calculations, Monkhorst-Pack k-meshes of about $0.025\times0.025\times0.025$ per Angstrom is adopted. Based on the DFT calculations, the maximally localized Wannier functions (MLWFs) \cite{cite20,cite21,cite22} are then constructed by using Wannier90 code \cite{cite23,cite24}, where a set of trial orbitals, $sp^3d^2$, $d_{xy}$, $d_{xz}$ and $d_{yz}$ are used for the initial guess for MLWFs.

Based on the MLWFs, the electronic conductivity can be calculated by employing the BTE method \cite{cite25,cite26,cite27,cite28}, where the chemical potential $\mu$ and temperature $T$ dependence of electronic conductivity can be obtained by
\begin{eqnarray}
\sigma _{ij}\left( \mu ,T \right) =e^2\int_{-\infty}^{+\infty}{d\varepsilon \left( -\frac{\partial f\left( \varepsilon ,\mu ,T \right)}{\partial \varepsilon} \right) \Sigma_{ij}\left(\varepsilon\right)}
	\label{eq1}
\end{eqnarray}
where $f\left( \varepsilon ,\mu ,T \right)$ is the Fermi-Dirac distribution function
\begin{eqnarray}
	f\left( \varepsilon ,\mu ,T \right)=\frac{1}{e^{\left(\varepsilon-\mu\right)/k_{\bm{B}}T}+1}
	\label{eq2}
\end{eqnarray}
and $\Sigma_{ij}\left(\varepsilon\right)$ is the transport distribution function (TDF) tensor defined as
\begin{eqnarray}
	\Sigma_{ij}\left(\varepsilon\right)=\frac{1}{V}\sum_{n,\bf{k}}{v_i\left(n,\bf{k}\right)v_j\left(n,\mathbf{k}\right)\tau\left(n,\bf{k}\right)\delta\left(\varepsilon-E_{n,\bf{k}}\right)}.
	\label{eq3}
\end{eqnarray}
Note that the sum in the above formulas is over all the energy bands (indexed by \textit{n}) with all states \textbf{k} (including spin, even if the spin index is not explicitly written here). $E_{n,\mathbf{k}}$ is the energy level and $v_i\left(n,\mathbf{k}\right)$ is the \textit{i}-th component of group velocity of the \textit{n}-th band in state \textbf{k}, $\delta$ is the Dirac’s delta function, $V=N_k\Omega_c$ corresponds to the total volume of the system, and $\tau$ is the relaxation time which is usually assumed to be a constant. In this study, we adopt $\tau=1.0\times 10^{-14}\,$s according to Ref. \cite{cite29}. In addition, a dense $\mathbf{k}$ mesh of 200 × 200 × 200 is employed to perform the Brillouin zone (BZ) integration for the electronic conductivity calculation.

\section{\label{sec:level3}result}

To make a direct comparison with the experimental observations, we focus on the electric current along [110] direction. Accordingly, the electronic conductivity $\sigma_{xx}$ and thus resistivity $\rho_{xx}=1/\sigma_{xx}$ is calculated with magnetization along $\left[\bar{1}10\right]$ and [001] direction, respectively. The structure details as well as schematic diagram for the Cartesian coordinates are shown in Fig. 1.

We first explore the influence of magnetization direction on the electronic structures without SOI. As shown in Figs. 2(a)–2(c), although the energy bands of spin-up and spin-down electrons significantly split along the electric current direction, they are independent on the magnetization direction. It is noted that the electronic conductance $\sigma_{xx}$[Eqs. (1)-(3)] is a function of electron velocity, Dirac’s delta function, and electron relaxation time, where the first two quantities depend on the band structure while the last one does not. Thus, the independence of band structure on magnetization direction means that no intrinsic UAMR exists in the absence of SOI.

\begin{figure}
	\centering
	\includegraphics[width=0.45\textwidth,height=0.32\textwidth]{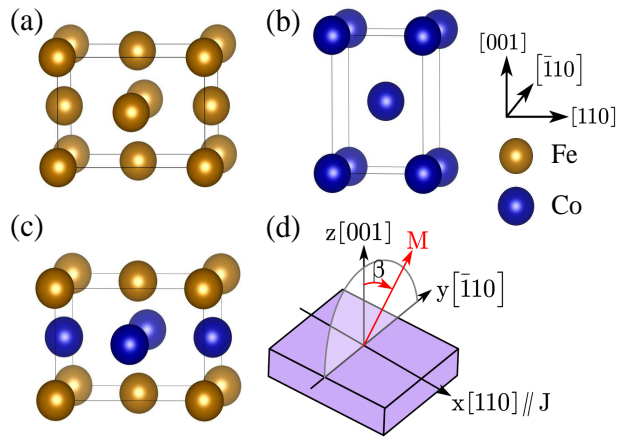}
	\caption{\label{fig1}Atomic structures of magnetic crystals (a-c) and schematic diagram for the definition of angle $\beta$ in the Cartesian coordinates (d). (a) bcc Fe (110), (b) hcp Co $\left(11\bar{2}0\right)$,  (c) bcc FeCo alloy (110), the brown and blue balls represent Fe and Co atoms, respectively. (d) Cartesian coordinates of sample. Electric current is along [110] direction defined as \textit{x} axis. $\left[\bar{1}10\right]$ and [001] are \textit{y}- and \textit{z}-axis respectively. Magnetization \textbf{M} is in \textit{yz}-plane.}
\end{figure}

\begin{figure*}
	\centering
	\includegraphics[width=0.9\textwidth,height=0.6\textwidth]{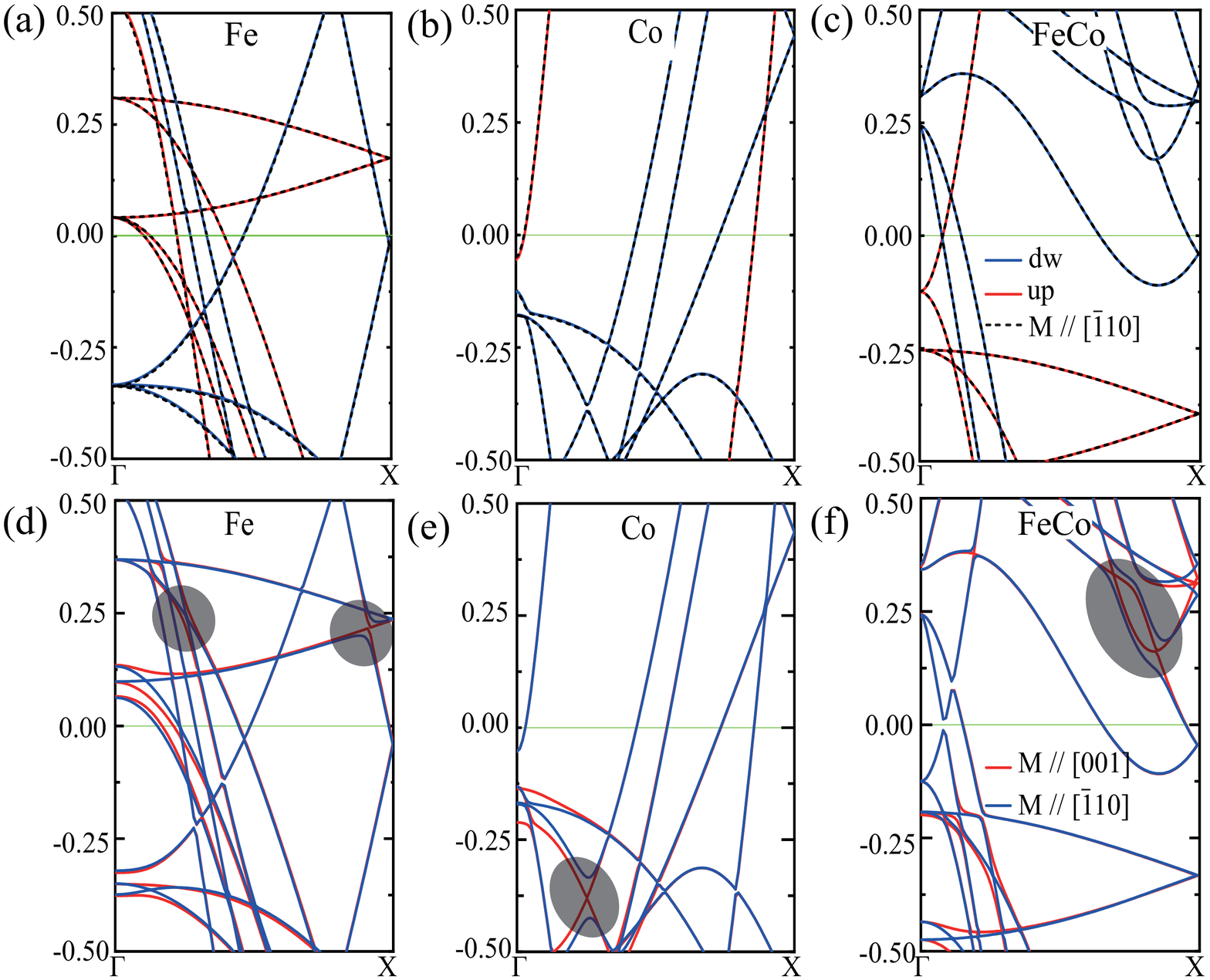}
	\caption{\label{fig2}Energy band without/with SOI for bcc Fe (110), hcp Co $\left(11\bar{2}0\right)$, and bcc FeCo (110) alloy when the magnetization is along [001] direction or $\left[\bar{1}10\right]$ direction, respectively. (a-c) Energy bands of the spin projection when the magnetization is along [001] direction and $\left[\bar{1}10\right]$ direction without SOI. The red and blue solid lines denote the energy bands of spin-up and spin-down electrons when the magnetization is along [001], respectively. The dashed lines are the energy bands when the magnetization is along  $\left[\bar{1}10\right]$. The solid and dashed lines completely coincide, showing the independence of energy bands on the magnetization direction without SOI. (d-f) Energy bands in the presence the SOI. The $\Gamma$-X in the BZ corresponds the x direction in the real space. Red and blue are the energy bands of spin-up and spin-down electrons in (a-c), and the bands of magnetization along the [001] direction and  $\left[\bar{1}10\right]$ direction in (d-f). The split of energy bands induced by different magnetization direction are indicated by the grey circles.}
\end{figure*}

Next, we discuss the magnetization-dependent SOI that affects band structures. Figs. 2(d)-2(f) show the energy bands of different magnetizations with SOI. It is clear that SOI induces band split in all three magnetic crystals. Notably, the band split strongly depends on the magnetization direction, and band structures are different when it is along [001] and $\left[\bar{1}10\right]$ directions. Particularly, the split becomes prominent for the intersected bands around certain \textbf{k} points. Since the change of band structure leads to the change of electron velocity and thus $\sigma_{xx}$, this result shows that the magnetization-dependent split of energy bands induced by SOI is responsible for the intrinsic UAMR. It is noticed that the dominate band split occurs in certain energy range, [0.15eV, 0.35eV] for bcc Fe (110) and bcc FeCo (110), and [-0.40eV, -0.25eV] for hcp Co $\left(11\bar{2}0\right)$ [marked by the grey circles in Figs. 2(d)-2(f)], which depends on the crystal structures. This feature indicates that the bcc and hcp structures have distinct UAMR behaviors.
 
\begin{figure*}
	\centering
	\includegraphics[width=0.9\textwidth,height=0.6\textwidth]{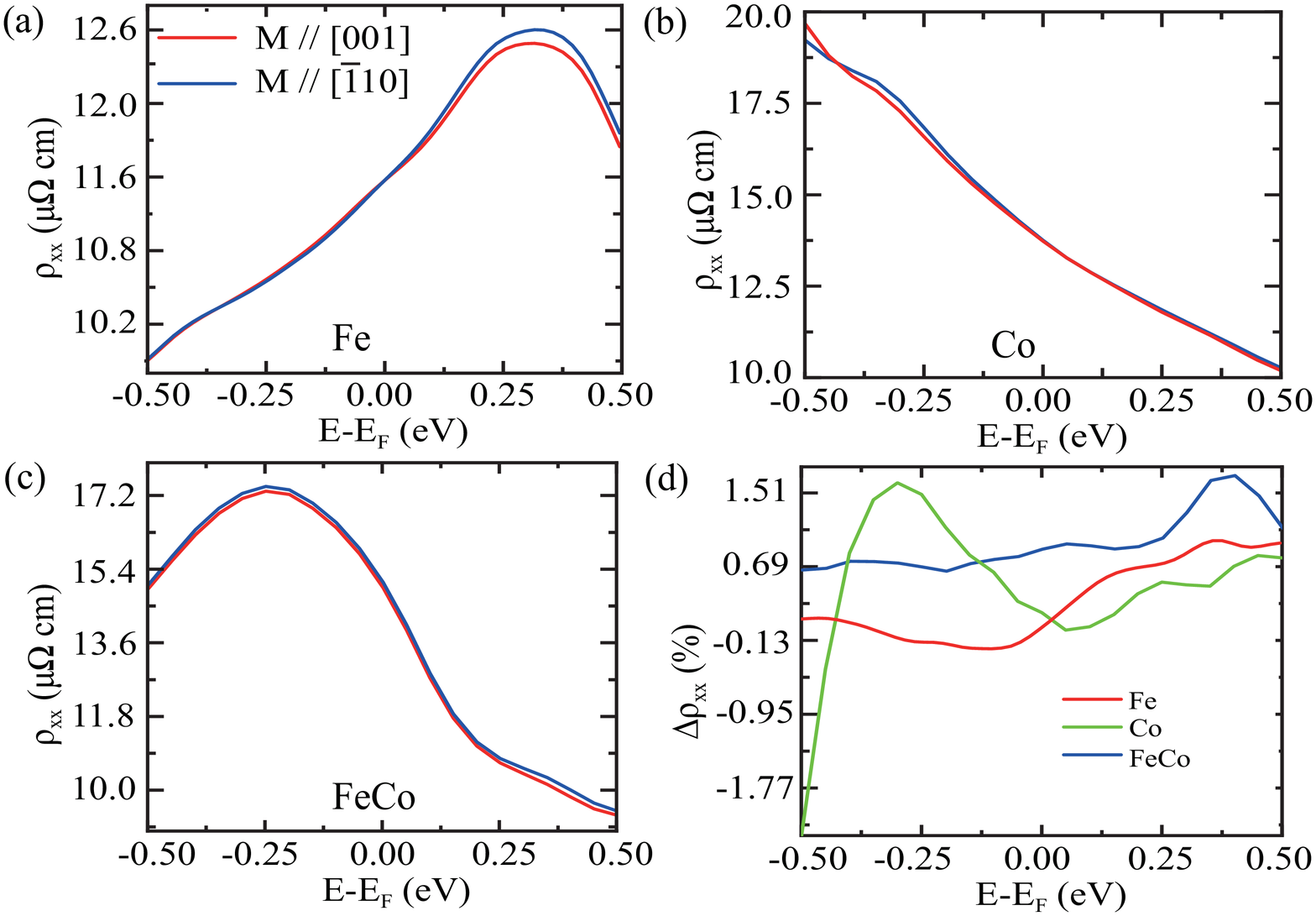}
	\caption{\label{fig3}Resistivity corresponding to different magnetization directions for (a) bcc Fe (110), (b) hcp Co $\left(11\bar{2}0\right)$, and (c) bcc FeCo (110) alloy, respectively. The red (blue) solid line represents the magnetization along the [110] ($\left[\bar{1}10\right]$) direction. (d) $\Delta\rho_{xx}$ varying with the position of Fermi energy of different materials.}
\end{figure*}

To confirm above analysis, we further calculate the electronic resistivity of the three magnetic crystals in the presence of the SOI. Figures 3(a)-3(c) show the energy level (relative to undoped crystal Fermi energy $E_F$) dependence of $\rho_{xx}$ for magnetization along [001] and $\left[\bar{1}10\right]$, respectively. It is found that $\rho_{xx}$ near $E_F$ for all the three materials are in the range of 11 - 15$\, \mu \Omega\ cm$, agree well with the experimental values around 10.8$\, \mu \Omega\ cm$ \cite{cite13}. Moreover, $\rho_{xx}$ of different magnetization directions differ from each other slightly, and the variations have obvious energy-dependence in all three magnetic crystals. Significant difference of $\rho_{xx}$ for magnetization along \textit{y}- and \textit{z}- directions occur around certain energy level with sizeable SOI induced band split. It is revealed by the following quantity, $\Delta\rho_{xx}=\left(\rho_{xx}^{\bar{1}10}-\rho_{xx}^{001}\right)/\rho_{xx}^{001}\times100\%$,
with $\rho_{xx}^{\bar{1}10}$ and $\rho_{xx}^{001}$  being the electronic resistivity for magnetization along $\left[\bar{1}10\right]$ and [001], respectively. As shown in Fig. 3(d), the maximum value of $\Delta\rho_{xx}$ in both bcc Fe (1.0\%) and bcc FeCo (1.6\%) appears around 0.37 eV above $E_F$, whereas $\Delta\rho_{xx}$ becomes the maximum around 0.3 eV below  $E_F$ for hcp Co (1.6\%). A common feature is that all the maximum value of $\Delta\rho_{xx}$ is 1.0\%-1.6\%, indicating the robustness of UAMR value which hardly depends on the component of Fe/Co in the FeCo alloy. The calculated $\Delta\rho_{xx}$ is in good agreement with the experimental value (1.0\%) \cite{cite13}, showing the reliability of BTE method for the UAMR. Note that the energy levels with the maximum $\Delta\rho_{xx}$ exactly locate in the energy range with the most significant band splits [Figs. 2(d)-2(f)]. These results clearly show the intrinsic UAMR originates from magnetization-dependent split of energy bands induced by SOI.

Then we reveal the intrinsic connection between SOI and order-parameters, i.e., magnetization and crystalline axis. According to the general form \cite{cite30}, the SOI for a conduction electron can be expressed as
\begin{eqnarray}
	H_{SOI}=\frac{\hbar e}{2mc^2}\left(\vec{\sigma}\times\vec{\nabla}U\right)\cdot\vec{v}_c
	\label{eq4}
\end{eqnarray}
where $U$ is the electrical potential and $\vec{\nabla}U$ corresponds to effective electrical field, $\vec{\sigma}$ is the spin, and $\vec{v}_c$ is the velocity of conduction electrons influenced by the external electrical field. Considering that the spin polarization of most conduction electrons in a magnetic metal is parallel to the magnetization direction $\vec{M}$, the magnetization-dependent SOI for the conduction electrons can be further obtained
\begin{eqnarray}
	H_{SOI}^M=\frac{\alpha\hbar e}{2mc^2}\left(\vec{M}\times\vec{\nabla}U\right)\cdot\vec{v}_c
	\label{eq5}
\end{eqnarray}
by defining $\vec{\sigma}=\alpha\vec{M}$, where $\alpha$ is a coefficient. Note that Eq.(5) works only for the conduction electrons with $\vec{\sigma}/\!/\vec{M}$.

On the other hand, as discussed in the supplemental materials to Ref. \cite{cite31}, $\vec{\nabla}U$ strongly depends on the spin-polarized charge density (SPCD) in the magnetic crystal. Considering that the majority-spin (denoted as $\uparrow$) electrons are in charge of the electronic conductivity, one can focus on SOI for the $\uparrow$ spin, where the effective electrical field can be expressed as $\vec{\nabla}U^{\uparrow}(r)\propto n^{\uparrow}(r)\vec{\nabla}n^{\uparrow}(r)$ with $n^{\uparrow}(r)$ and $\vec{\nabla}n^{\uparrow}(r)$ being the SPCD and its gradient at position $\vec{r}$ for the $\uparrow$ electrons, respectively \cite{cite31}. Hence, it is obvious that different $\vec{\nabla}U^{\uparrow}(r)$ values would be obtained if only $n^{\uparrow}(r)$ and $\vec{\nabla}n^{\uparrow}(r)$ are anisotropic in a crystal. In fact, such requirement can be easily satisfied. For example, although bcc Fe has very high geometric symmetry, its $n^{\uparrow}(r)$ [$\vec{\nabla}n^{\uparrow}(r)$] along $\left[\bar{1}10\right]$ and [001] directions are still distinct according to the symmetry analysis, which results in $\nabla_{\bar{1}10}U^{\uparrow}\ne\nabla_{001}U^{\uparrow}$. According to Eq. (5), the variation of SOI$^{\prime}s$ strength is attributed by $\vec{M}\times\vec{\nabla}U$ when $\vec{v}_c$ is fixed, where the strongest SOI corresponds to the case $\vec{M}\perp\vec{\nabla}U$. Therefore, although the magnetization\ is perpendicular to the electric current (along [110] direction) in both cases of $\vec{M}$/\!/$\left[001\right]$ and $\vec{M}$/\!/$\left[\bar{1}10\right]$, different SOI strength can be still induced by the fact that $\nabla_{\bar{1}10}U^{\uparrow}\ne\nabla_{001}U^{\uparrow}$. Note that such mechanism should be general, which also applies to hcp Co, and bcc FeCo.

\begin{figure*}
	\centering
	\includegraphics[width=0.9\textwidth,height=0.35\textwidth]{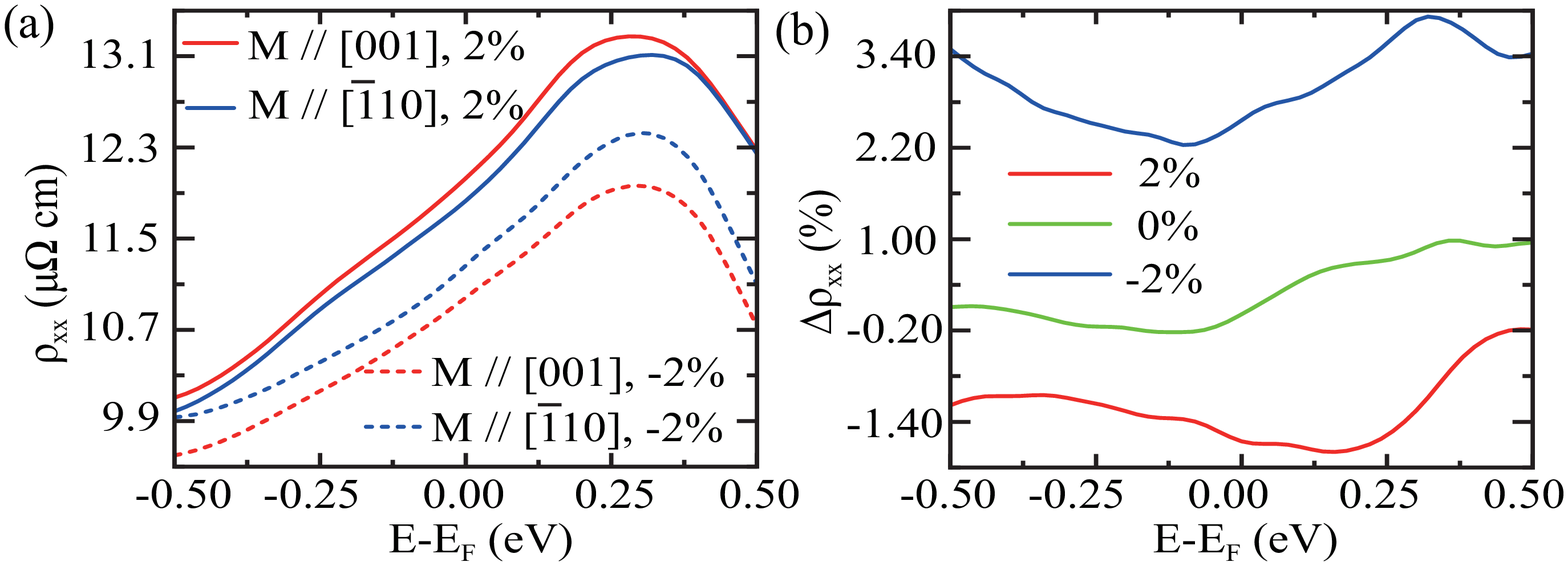}
	\caption{\label{fig4}(a) Resistivity of bcc Fe (110) corresponding to different magnetization directions at strain of -2\% and 2\%. Red (blue) solid and dashed lines represent the $\rho_{xx}$ when the magnetization is along the [110] $(\left[\bar{1}10\right])$ direction, and the strain is applied at 2\% and -2\%, respectively. (b) $\Delta\rho_{xx}$ of Fe (110) varying with the position of Fermi energy under different strains.}
\end{figure*}

The above results clearly reveal the intrinsic correlation among magnetization (corresponds to $\vec{M}$), crystalline axis (corresponds to $\vec{\nabla}U$), and external electrical field (corresponds to $\vec{v_c})$. The coupling between order parameters, i.e., magnetization and srystal axis,  leads to the magnetization-dependent SOI and thus UAMR. This mechanism is different from the one in conventional AMR, which solely comes from the coupling between magnetization and external electrical field (or electric current) and ignores the crystalline-axis dependence of  $\vec{\nabla}U$ \cite{cite32,cite33,cite34}.

To confirm above results, we further investigate the biaxially strain-dependent UAMR. The biaxial strain is expected to change the anisotropy of SPCD and thus the difference between $\nabla_{\bar{1}10}U^{\uparrow}\ne\nabla_{001}U^{\uparrow}$, which results in significant variation of UAMR. Figures 4(a) shows the energy level (relative to $E_F$) dependence of $\rho_{xx}$ of bcc Fe with the magnetization along [001] and $[\bar{1}10]$ at different strains, respectively. It is found that the compressive and tensile strains have completely different effect on $\rho_{xx}$, which leads to the divergence of UAMR. When a compressive strain (-2\%) is applied, $\rho_{xx}$ with magnetization along $[\bar{1}10]$ is significantly larger than that along [001] in the whole energy region near $E_F$. Whereas it has a converse behavior when a tensile strain (2\%) is applied. The corresponding $\Delta\rho_{xx}$ is additionally shown in Fig. 4(b). As one can see, strain can significantly enhance UAMR, the maximum $|\Delta\rho_{xx}|$ reaches 4.0 and 1.6 times larger than that without strain (1.0\%) under the compressive and tensile strains, respectively. This result confirms the intrinsic coupling between magnetization and crystalline lattice, which gives rise to the magnetization-dependent SOI and thus UAMR.

Before conclusion, it should be pointed out that present UAMR theory is for the homogeneous materials, 
not for bilayers \cite{XRW}.

\section{\label{sec:level4}conclusion}

In summary, we find UAMR in Fe, Co and FeCo alloy originated from the magnetization-dependent SOI. Using the DFT calculations and BTE we show that SOI splits intersecting energy bands around the Fermi level. Different magnetization direction gives rise to different SOIs that, in turn, lead to different resistivity, or UAMR. This is an intrinsic UAMR. SOI is essential in UAMR since the energy bands do not depend on magnetization direction in the absence of the SOI. A phenomenological model that reveals the relationship between SOI and order-parameters (magnetization and crystalline lattice) is proposed and is confirmed by the strong strain dependence of UAMR.



\begin{thebibliography}{35}%
	\makeatletter
	\providecommand \@ifxundefined [1]{%
		\@ifx{#1\undefined}
	}%
	\providecommand \@ifnum [1]{%
		\ifnum #1\expandafter \@firstoftwo
		\else \expandafter \@secondoftwo
		\fi
	}%
	\providecommand \@ifx [1]{%
		\ifx #1\expandafter \@firstoftwo
		\else \expandafter \@secondoftwo
		\fi
	}%
	\providecommand \natexlab [1]{#1}%
	\providecommand \enquote  [1]{``#1''}%
	\providecommand \bibnamefont  [1]{#1}%
	\providecommand \bibfnamefont [1]{#1}%
	\providecommand \citenamefont [1]{#1}%
	\providecommand \href@noop [0]{\@secondoftwo}%
	\providecommand \href [0]{\begingroup \@sanitize@url \@href}%
	\providecommand \@href[1]{\@@startlink{#1}\@@href}%
	\providecommand \@@href[1]{\endgroup#1\@@endlink}%
	\providecommand \@sanitize@url [0]{\catcode `\\12\catcode `\$12\catcode
		`\&12\catcode `\#12\catcode `\^12\catcode `\_12\catcode `\%12\relax}%
	\providecommand \@@startlink[1]{}%
	\providecommand \@@endlink[0]{}%
	\providecommand \url  [0]{\begingroup\@sanitize@url \@url }%
	\providecommand \@url [1]{\endgroup\@href {#1}{\urlprefix }}%
	\providecommand \urlprefix  [0]{URL }%
	\providecommand \Eprint [0]{\href }%
	\providecommand \doibase [0]{https://doi.org/}%
	\providecommand \selectlanguage [0]{\@gobble}%
	\providecommand \bibinfo  [0]{\@secondoftwo}%
	\providecommand \bibfield  [0]{\@secondoftwo}%
	\providecommand \translation [1]{[#1]}%
	\providecommand \BibitemOpen [0]{}%
	\providecommand \bibitemStop [0]{}%
	\providecommand \bibitemNoStop [0]{.\EOS\space}%
	\providecommand \EOS [0]{\spacefactor3000\relax}%
	\providecommand \BibitemShut  [1]{\csname bibitem#1\endcsname}%
	\let\auto@bib@innerbib\@empty
	\bibitem [{\citenamefont {Chappert}\ \emph {et~al.}(2007)\citenamefont
		{Chappert}, \citenamefont {Fert},\ and\ \citenamefont {Van~Dau}}]{cite1}%
	\BibitemOpen
	\bibfield  {author} {\bibinfo {author} {\bibfnamefont {C.}~\bibnamefont
			{Chappert}}, \bibinfo {author} {\bibfnamefont {A.}~\bibnamefont {Fert}},\
		and\ \bibinfo {author} {\bibfnamefont {F.~N.}\ \bibnamefont {Van~Dau}},\
	}\bibfield  {title} {\bibinfo {title} {{The Emergence of Spin Electronics in
				Data Storage}},\ }\href@noop {} {\bibfield  {journal} {\bibinfo  {journal}
			{Nat. Mater.}\ }\textbf {\bibinfo {volume} {6}},\ \bibinfo {pages} {813}
		(\bibinfo {year} {2007})}\BibitemShut {NoStop}%
	\bibitem [{\citenamefont {D{\"o}ring}(1938)}]{cite2}%
	\BibitemOpen
	\bibfield  {author} {\bibinfo {author} {\bibfnamefont {W.}~\bibnamefont
			{D{\"o}ring}},\ }\bibfield  {title} {\bibinfo {title} {{Die Abh\"angigkeit
				Des Widerstandes von Nickelkristallen von der Richtung der spontanen
				Magnetisierung}},\ }\href@noop {} {\bibfield  {journal} {\bibinfo  {journal}
			{Ann. Phys.}\ }\textbf {\bibinfo {volume} {424}},\ \bibinfo {pages} {259}
		(\bibinfo {year} {1938})}\BibitemShut {NoStop}%
	\bibitem [{\citenamefont {McGuire}\ and\ \citenamefont {Potter}(1975)}]{cite3}%
	\BibitemOpen
	\bibfield  {author} {\bibinfo {author} {\bibfnamefont {T.}~\bibnamefont
			{McGuire}}\ and\ \bibinfo {author} {\bibfnamefont {R.}~\bibnamefont
			{Potter}},\ }\bibfield  {title} {\bibinfo {title} {{Anisotropic
				Magnetoresistance in Ferromagnetic 3d Alloys}},\ }\href@noop {} {\bibfield
		{journal} {\bibinfo  {journal} {IEEE Trans. Magn.}\ }\textbf {\bibinfo
			{volume} {11}},\ \bibinfo {pages} {1018} (\bibinfo {year}
		{1975})}\BibitemShut {NoStop}%
	\bibitem [{\citenamefont {Berger}(1964)}]{cite4}%
	\BibitemOpen
	\bibfield  {author} {\bibinfo {author} {\bibfnamefont {L.}~\bibnamefont
			{Berger}},\ }\bibfield  {title} {\bibinfo {title} {{Influence of Spin-Orbit
				Interaction on the Transport Processes in Ferromagnetic Nickel Alloys, in the
				Presence of a Degeneracy of the 3d Band}},\ }\href@noop {} {\bibfield
		{journal} {\bibinfo  {journal} {Physica}\ }\textbf {\bibinfo {volume} {30}},\
		\bibinfo {pages} {1141} (\bibinfo {year} {1964})}\BibitemShut {NoStop}%
	\bibitem [{\citenamefont {Campbell}\ \emph
		{et~al.}(1970{\natexlab{a}})\citenamefont {Campbell}, \citenamefont {Fert},\
		and\ \citenamefont {Jaoul}}]{cite5}%
	\BibitemOpen
	\bibfield  {author} {\bibinfo {author} {\bibfnamefont {I.~A.}\ \bibnamefont
			{Campbell}}, \bibinfo {author} {\bibfnamefont {A.}~\bibnamefont {Fert}},\
		and\ \bibinfo {author} {\bibfnamefont {O.}~\bibnamefont {Jaoul}},\ }\bibfield
	{title} {\bibinfo {title} {{The Spontaneous Resistivity Anisotropy in
				Ni-Based Alloys}},\ }\href@noop {} {\bibfield  {journal} {\bibinfo  {journal}
			{J. Phys. C: Solid State Phys.}\ }\textbf {\bibinfo {volume} {3}},\ \bibinfo
		{pages} {S95} (\bibinfo {year} {1970}{\natexlab{a}})}\BibitemShut {NoStop}%
	\bibitem [{\citenamefont {Seemann}\ \emph {et~al.}(2011)\citenamefont
		{Seemann}, \citenamefont {Freimuth}, \citenamefont {Zhang}, \citenamefont
		{Bl{\"u}gel}, \citenamefont {Mokrousov}, \citenamefont {B{\"u}rgler},\ and\
		\citenamefont {Schneider}}]{cite6}%
	\BibitemOpen
	\bibfield  {author} {\bibinfo {author} {\bibfnamefont {K.~M.}\ \bibnamefont
			{Seemann}}, \bibinfo {author} {\bibfnamefont {F.}~\bibnamefont {Freimuth}},
		\bibinfo {author} {\bibfnamefont {H.}~\bibnamefont {Zhang}}, \bibinfo
		{author} {\bibfnamefont {S.}~\bibnamefont {Bl{\"u}gel}}, \bibinfo {author}
		{\bibfnamefont {Y.}~\bibnamefont {Mokrousov}}, \bibinfo {author}
		{\bibfnamefont {D.~E.}\ \bibnamefont {B{\"u}rgler}},\ and\ \bibinfo {author}
		{\bibfnamefont {C.~M.}\ \bibnamefont {Schneider}},\ }\bibfield  {title}
	{\bibinfo {title} {{Origin of the Planar Hall Effect in Nanocrystalline
				Co\textsubscript{60}Fe\textsubscript{20}B\textsubscript{20}}},\ }\href@noop
	{} {\bibfield  {journal} {\bibinfo  {journal} {Phys. Rev. Lett.}\ }\textbf
		{\bibinfo {volume} {107}},\ \bibinfo {pages} {086603} (\bibinfo {year}
		{2011})}\BibitemShut {NoStop}%
	\bibitem [{\citenamefont {N{\'a}dvorn{\'i}k}\ \emph {et~al.}(2021)\citenamefont
		{N{\'a}dvorn{\'i}k}, \citenamefont {Borchert}, \citenamefont {Brandt},
		\citenamefont {Schlitz}, \citenamefont {{de Mare}}, \citenamefont
		{V{\'y}born{\'y}}, \citenamefont {Mertig}, \citenamefont {Jakob},
		\citenamefont {Kl{\"a}ui}, \citenamefont {Goennenwein}, \citenamefont {Wolf},
		\citenamefont {Woltersdorf},\ and\ \citenamefont {Kampfrath}}]{cite7}%
	\BibitemOpen
	\bibfield  {author} {\bibinfo {author} {\bibfnamefont {L.}~\bibnamefont
			{N{\'a}dvorn{\'i}k}}, \bibinfo {author} {\bibfnamefont {M.}~\bibnamefont
			{Borchert}}, \bibinfo {author} {\bibfnamefont {L.}~\bibnamefont {Brandt}},
		\bibinfo {author} {\bibfnamefont {R.}~\bibnamefont {Schlitz}}, \bibinfo
		{author} {\bibfnamefont {K.~A.}\ \bibnamefont {{de Mare}}}, \bibinfo {author}
		{\bibfnamefont {K.}~\bibnamefont {V{\'y}born{\'y}}}, \bibinfo {author}
		{\bibfnamefont {I.}~\bibnamefont {Mertig}}, \bibinfo {author} {\bibfnamefont
			{G.}~\bibnamefont {Jakob}}, \bibinfo {author} {\bibfnamefont
			{M.}~\bibnamefont {Kl{\"a}ui}}, \bibinfo {author} {\bibfnamefont {S.~T.~B.}\
			\bibnamefont {Goennenwein}}, \bibinfo {author} {\bibfnamefont
			{M.}~\bibnamefont {Wolf}}, \bibinfo {author} {\bibfnamefont {G.}~\bibnamefont
			{Woltersdorf}},\ and\ \bibinfo {author} {\bibfnamefont {T.}~\bibnamefont
			{Kampfrath}},\ }\bibfield  {title} {\bibinfo {title} {{Broadband Terahertz
				Probes of Anisotropic Magnetoresistance Disentangle Extrinsic and Intrinsic
				Contributions}},\ }\href@noop {} {\bibfield  {journal} {\bibinfo  {journal}
			{Phys. Rev. X}\ }\textbf {\bibinfo {volume} {11}},\ \bibinfo {pages} {021030}
		(\bibinfo {year} {2021})}\BibitemShut {NoStop}%
	\bibitem [{\citenamefont {Tomar}\ \emph {et~al.}(2021)\citenamefont {Tomar},
		\citenamefont {Kakkar}, \citenamefont {Bera},\ and\ \citenamefont
		{Chakraverty}}]{cite8}%
	\BibitemOpen
	\bibfield  {author} {\bibinfo {author} {\bibfnamefont {R.}~\bibnamefont
			{Tomar}}, \bibinfo {author} {\bibfnamefont {S.}~\bibnamefont {Kakkar}},
		\bibinfo {author} {\bibfnamefont {C.}~\bibnamefont {Bera}},\ and\ \bibinfo
		{author} {\bibfnamefont {S.}~\bibnamefont {Chakraverty}},\ }\bibfield
	{title} {\bibinfo {title} {{Anisotropic Magnetoresistance and Planar Hall
				Effect in (001) and (111) LaVO\textsubscript{3}/SrTiO\textsubscript{3}
				Heterostructures}},\ }\href@noop {} {\bibfield  {journal} {\bibinfo
			{journal} {Phys. Rev. B}\ }\textbf {\bibinfo {volume} {103}},\ \bibinfo
		{pages} {115407} (\bibinfo {year} {2021})}\BibitemShut {NoStop}%
	\bibitem [{\citenamefont {Miao}\ \emph {et~al.}(2021)\citenamefont {Miao},
		\citenamefont {Yang}, \citenamefont {Jia}, \citenamefont {Li}, \citenamefont
		{Yang}, \citenamefont {Gao},\ and\ \citenamefont {Xue}}]{cite9}%
	\BibitemOpen
	\bibfield  {author} {\bibinfo {author} {\bibfnamefont {Y.}~\bibnamefont
			{Miao}}, \bibinfo {author} {\bibfnamefont {D.}~\bibnamefont {Yang}}, \bibinfo
		{author} {\bibfnamefont {L.}~\bibnamefont {Jia}}, \bibinfo {author}
		{\bibfnamefont {X.}~\bibnamefont {Li}}, \bibinfo {author} {\bibfnamefont
			{S.}~\bibnamefont {Yang}}, \bibinfo {author} {\bibfnamefont {C.}~\bibnamefont
			{Gao}},\ and\ \bibinfo {author} {\bibfnamefont {D.}~\bibnamefont {Xue}},\
	}\bibfield  {title} {\bibinfo {title} {{Magnetocrystalline Anisotropy
				Correlated Negative Anisotropic Magnetoresistance in Epitaxial
				Fe\textsubscript{30}Co\textsubscript{70} Thin Films}},\ }\href@noop {}
	{\bibfield  {journal} {\bibinfo  {journal} {Appl. Phys. Lett.}\ }\textbf
		{\bibinfo {volume} {118}},\ \bibinfo {pages} {042404} (\bibinfo {year}
		{2021})}\BibitemShut {NoStop}%
	\bibitem [{\citenamefont {Ritzinger}\ \emph {et~al.}(2021)\citenamefont
		{Ritzinger}, \citenamefont {Reichlova}, \citenamefont {Kriegner},
		\citenamefont {Markou}, \citenamefont {Schlitz}, \citenamefont {Lammel},
		\citenamefont {Scheffler}, \citenamefont {Park}, \citenamefont {Thomas},
		\citenamefont {St{\v r}eda}, \citenamefont {Felser}, \citenamefont
		{Goennenwein},\ and\ \citenamefont {V{\'y}born{\'y}}}]{cite10}%
	\BibitemOpen
	\bibfield  {author} {\bibinfo {author} {\bibfnamefont {P.}~\bibnamefont
			{Ritzinger}}, \bibinfo {author} {\bibfnamefont {H.}~\bibnamefont
			{Reichlova}}, \bibinfo {author} {\bibfnamefont {D.}~\bibnamefont {Kriegner}},
		\bibinfo {author} {\bibfnamefont {A.}~\bibnamefont {Markou}}, \bibinfo
		{author} {\bibfnamefont {R.}~\bibnamefont {Schlitz}}, \bibinfo {author}
		{\bibfnamefont {M.}~\bibnamefont {Lammel}}, \bibinfo {author} {\bibfnamefont
			{D.}~\bibnamefont {Scheffler}}, \bibinfo {author} {\bibfnamefont {G.~H.}\
			\bibnamefont {Park}}, \bibinfo {author} {\bibfnamefont {A.}~\bibnamefont
			{Thomas}}, \bibinfo {author} {\bibfnamefont {P.}~\bibnamefont {St{\v r}eda}},
		\bibinfo {author} {\bibfnamefont {C.}~\bibnamefont {Felser}}, \bibinfo
		{author} {\bibfnamefont {S.~T.~B.}\ \bibnamefont {Goennenwein}},\ and\
		\bibinfo {author} {\bibfnamefont {K.}~\bibnamefont {V{\'y}born{\'y}}},\
	}\bibfield  {title} {\bibinfo {title} {{Anisotropic Magnetothermal Transport
				in Co\textsubscript{2}MnGa Thin Films}},\ }\href@noop {} {\bibfield
		{journal} {\bibinfo  {journal} {Phys. Rev. B}\ }\textbf {\bibinfo {volume}
			{104}},\ \bibinfo {pages} {094406} (\bibinfo {year} {2021})}\BibitemShut
	{NoStop}%
	\bibitem [{\citenamefont {Zeng}\ \emph
		{et~al.}(2020{\natexlab{a}})\citenamefont {Zeng}, \citenamefont {Zhou},
		\citenamefont {Jia}, \citenamefont {Shi}, \citenamefont {Huo}, \citenamefont
		{Zhang},\ and\ \citenamefont {Wu}}]{cite11}%
	\BibitemOpen
	\bibfield  {author} {\bibinfo {author} {\bibfnamefont {F.}~\bibnamefont
			{Zeng}}, \bibinfo {author} {\bibfnamefont {C.}~\bibnamefont {Zhou}}, \bibinfo
		{author} {\bibfnamefont {M.}~\bibnamefont {Jia}}, \bibinfo {author}
		{\bibfnamefont {D.}~\bibnamefont {Shi}}, \bibinfo {author} {\bibfnamefont
			{Y.}~\bibnamefont {Huo}}, \bibinfo {author} {\bibfnamefont {W.}~\bibnamefont
			{Zhang}},\ and\ \bibinfo {author} {\bibfnamefont {Y.}~\bibnamefont {Wu}},\
	}\bibfield  {title} {\bibinfo {title} {{Strong Current-Direction Dependence
				of Anisotropic Magnetoresistance in Single Crystalline Fe/GaAs(110) Films}},\
	}\href@noop {} {\bibfield  {journal} {\bibinfo  {journal} {J. Magn. Magn.
				Mater.}\ }\textbf {\bibinfo {volume} {499}},\ \bibinfo {pages} {166204}
		(\bibinfo {year} {2020}{\natexlab{a}})}\BibitemShut {NoStop}%
	\bibitem [{\citenamefont {Thompson}\ \emph {et~al.}(1975)\citenamefont
		{Thompson}, \citenamefont {Romankiw},\ and\ \citenamefont
		{Mayadas}}]{cite12}%
	\BibitemOpen
	\bibfield  {author} {\bibinfo {author} {\bibfnamefont {D.}~\bibnamefont
			{Thompson}}, \bibinfo {author} {\bibfnamefont {L.}~\bibnamefont {Romankiw}},\
		and\ \bibinfo {author} {\bibfnamefont {A.}~\bibnamefont {Mayadas}},\
	}\bibfield  {title} {\bibinfo {title} {{Thin film magnetoresistors in memory,
				storage, and related applications}},\ }\href
	{https://doi.org/10.1109/TMAG.1975.1058786} {\bibfield  {journal} {\bibinfo
			{journal} {IEEE Trans. Magn.}\ }\textbf {\bibinfo {volume} {11}},\ \bibinfo
		{pages} {1039} (\bibinfo {year} {1975})}\BibitemShut {NoStop}%
	\bibitem [{\citenamefont {Zeng}\ \emph
		{et~al.}(2020{\natexlab{b}})\citenamefont {Zeng}, \citenamefont {Ren},
		\citenamefont {Li}, \citenamefont {Zeng}, \citenamefont {Jia}, \citenamefont
		{Miao}, \citenamefont {Hoffmann}, \citenamefont {Zhang}, \citenamefont {Wu},\
		and\ \citenamefont {Yuan}}]{cite13}%
	\BibitemOpen
	\bibfield  {author} {\bibinfo {author} {\bibfnamefont {F.~L.}\ \bibnamefont
			{Zeng}}, \bibinfo {author} {\bibfnamefont {Z.~Y.}\ \bibnamefont {Ren}},
		\bibinfo {author} {\bibfnamefont {Y.}~\bibnamefont {Li}}, \bibinfo {author}
		{\bibfnamefont {J.~Y.}\ \bibnamefont {Zeng}}, \bibinfo {author}
		{\bibfnamefont {M.~W.}\ \bibnamefont {Jia}}, \bibinfo {author} {\bibfnamefont
			{J.}~\bibnamefont {Miao}}, \bibinfo {author} {\bibfnamefont {A.}~\bibnamefont
			{Hoffmann}}, \bibinfo {author} {\bibfnamefont {W.}~\bibnamefont {Zhang}},
		\bibinfo {author} {\bibfnamefont {Y.~Z.}\ \bibnamefont {Wu}},\ and\ \bibinfo
		{author} {\bibfnamefont {Z.}~\bibnamefont {Yuan}},\ }\bibfield  {title}
	{\bibinfo {title} {{Intrinsic Mechanism for Anisotropic Magnetoresistance and
				Experimental Confirmation in
				Co\textsubscript{\textit{x}}Fe\textsubscript{1-\textit{x}} Single-Crystal
				Films}},\ }\href@noop {} {\bibfield  {journal} {\bibinfo  {journal} {Phys.
				Rev. Lett.}\ }\textbf {\bibinfo {volume} {125}},\ \bibinfo {pages} {097201}
		(\bibinfo {year} {2020}{\natexlab{b}})}\BibitemShut {NoStop}%
	\bibitem [{\citenamefont {De~Ranieri}\ \emph {et~al.}(2008)\citenamefont
		{De~Ranieri}, \citenamefont {Rushforth}, \citenamefont {V{\'y}born{\'y}},
		\citenamefont {Rana}, \citenamefont {Ahmad}, \citenamefont {Campion},
		\citenamefont {Foxon}, \citenamefont {Gallagher}, \citenamefont {Irvine},
		\citenamefont {Wunderlich},\ and\ \citenamefont {Jungwirth}}]{cite14}%
	\BibitemOpen
	\bibfield  {author} {\bibinfo {author} {\bibfnamefont {E.}~\bibnamefont
			{De~Ranieri}}, \bibinfo {author} {\bibfnamefont {A.~W.}\ \bibnamefont
			{Rushforth}}, \bibinfo {author} {\bibfnamefont {K.}~\bibnamefont
			{V{\'y}born{\'y}}}, \bibinfo {author} {\bibfnamefont {U.}~\bibnamefont
			{Rana}}, \bibinfo {author} {\bibfnamefont {E.}~\bibnamefont {Ahmad}},
		\bibinfo {author} {\bibfnamefont {R.~P.}\ \bibnamefont {Campion}}, \bibinfo
		{author} {\bibfnamefont {C.~T.}\ \bibnamefont {Foxon}}, \bibinfo {author}
		{\bibfnamefont {B.~L.}\ \bibnamefont {Gallagher}}, \bibinfo {author}
		{\bibfnamefont {A.~C.}\ \bibnamefont {Irvine}}, \bibinfo {author}
		{\bibfnamefont {J.}~\bibnamefont {Wunderlich}},\ and\ \bibinfo {author}
		{\bibfnamefont {T.}~\bibnamefont {Jungwirth}},\ }\bibfield  {title} {\bibinfo
		{title} {{Lithographically and Electrically Controlled Strain Effects on
				Anisotropic Magnetoresistance in (Ga,Mn)As}},\ }\href@noop {} {\bibfield
		{journal} {\bibinfo  {journal} {New J. Phys.}\ }\textbf {\bibinfo {volume}
			{10}},\ \bibinfo {pages} {065003} (\bibinfo {year} {2008})}\BibitemShut
	{NoStop}%
	\bibitem [{\citenamefont {Zhang}\ \emph {et~al.}(2016)\citenamefont {Zhang},
		\citenamefont {Wang},\ and\ \citenamefont {Zhang}}]{cite15}%
	\BibitemOpen
	\bibfield  {author} {\bibinfo {author} {\bibfnamefont {Y.}~\bibnamefont
			{Zhang}}, \bibinfo {author} {\bibfnamefont {X.~R.}\ \bibnamefont {Wang}},\
		and\ \bibinfo {author} {\bibfnamefont {H.~W.}\ \bibnamefont {Zhang}},\
	}\bibfield  {title} {\bibinfo {title} {{Extraordinary Galvanomagnetic Effects
				in Polycrystalline Magnetic Films}},\ }\href@noop {} {\bibfield  {journal}
		{\bibinfo  {journal} {Europhys. Lett.}\ }\textbf {\bibinfo {volume} {113}},\
		\bibinfo {pages} {47003} (\bibinfo {year} {2016})}\BibitemShut {NoStop}%
	\bibitem [{\citenamefont {Wang}(2022)}]{cite16}%
	\BibitemOpen
	\bibfield  {author} {\bibinfo {author} {\bibfnamefont {X.~R.}\ \bibnamefont
			{Wang}},\ }\bibfield  {title} {\bibinfo {title} {{A Theory for Anisotropic
				Magnetoresistance in Materials with Two Vector Order Parameters}},\
	}\href@noop {} {\bibfield  {journal} {\bibinfo  {journal} {Chin. Phys.
				Lett.}\ }\textbf {\bibinfo {volume} {39}},\ \bibinfo {pages} {027301}
		(\bibinfo {year} {2022})}\BibitemShut {NoStop}%
	\bibitem [{\citenamefont {Giannozzi}\ \emph {et~al.}(2009)\citenamefont
		{Giannozzi}, \citenamefont {Baroni}, \citenamefont {Bonini}, \citenamefont
		{Calandra}, \citenamefont {Car}, \citenamefont {Cavazzoni}, \citenamefont
		{Ceresoli}, \citenamefont {Chiarotti}, \citenamefont {Cococcioni},
		\citenamefont {Dabo}, \citenamefont {Dal~Corso}, \citenamefont {{de
				Gironcoli}}, \citenamefont {Fabris}, \citenamefont {Fratesi}, \citenamefont
		{Gebauer}, \citenamefont {Gerstmann}, \citenamefont {Gougoussis},
		\citenamefont {Kokalj}, \citenamefont {Lazzeri}, \citenamefont
		{{Martin-Samos}}, \citenamefont {Marzari}, \citenamefont {Mauri},
		\citenamefont {Mazzarello}, \citenamefont {Paolini}, \citenamefont
		{Pasquarello}, \citenamefont {Paulatto}, \citenamefont {Sbraccia},
		\citenamefont {Scandolo}, \citenamefont {Sclauzero}, \citenamefont
		{Seitsonen}, \citenamefont {Smogunov}, \citenamefont {Umari},\ and\
		\citenamefont {Wentzcovitch}}]{cite17}%
	\BibitemOpen
	\bibfield  {author} {\bibinfo {author} {\bibfnamefont {P.}~\bibnamefont
			{Giannozzi}}, \bibinfo {author} {\bibfnamefont {S.}~\bibnamefont {Baroni}},
		\bibinfo {author} {\bibfnamefont {N.}~\bibnamefont {Bonini}}, \bibinfo
		{author} {\bibfnamefont {M.}~\bibnamefont {Calandra}}, \bibinfo {author}
		{\bibfnamefont {R.}~\bibnamefont {Car}}, \bibinfo {author} {\bibfnamefont
			{C.}~\bibnamefont {Cavazzoni}}, \bibinfo {author} {\bibfnamefont
			{D.}~\bibnamefont {Ceresoli}}, \bibinfo {author} {\bibfnamefont {G.~L.}\
			\bibnamefont {Chiarotti}}, \bibinfo {author} {\bibfnamefont {M.}~\bibnamefont
			{Cococcioni}}, \bibinfo {author} {\bibfnamefont {I.}~\bibnamefont {Dabo}},
		\bibinfo {author} {\bibfnamefont {A.}~\bibnamefont {Dal~Corso}}, \bibinfo
		{author} {\bibfnamefont {S.}~\bibnamefont {{de Gironcoli}}}, \bibinfo
		{author} {\bibfnamefont {S.}~\bibnamefont {Fabris}}, \bibinfo {author}
		{\bibfnamefont {G.}~\bibnamefont {Fratesi}}, \bibinfo {author} {\bibfnamefont
			{R.}~\bibnamefont {Gebauer}}, \bibinfo {author} {\bibfnamefont
			{U.}~\bibnamefont {Gerstmann}}, \bibinfo {author} {\bibfnamefont
			{C.}~\bibnamefont {Gougoussis}}, \bibinfo {author} {\bibfnamefont
			{A.}~\bibnamefont {Kokalj}}, \bibinfo {author} {\bibfnamefont
			{M.}~\bibnamefont {Lazzeri}}, \bibinfo {author} {\bibfnamefont
			{L.}~\bibnamefont {{Martin-Samos}}}, \bibinfo {author} {\bibfnamefont
			{N.}~\bibnamefont {Marzari}}, \bibinfo {author} {\bibfnamefont
			{F.}~\bibnamefont {Mauri}}, \bibinfo {author} {\bibfnamefont
			{R.}~\bibnamefont {Mazzarello}}, \bibinfo {author} {\bibfnamefont
			{S.}~\bibnamefont {Paolini}}, \bibinfo {author} {\bibfnamefont
			{A.}~\bibnamefont {Pasquarello}}, \bibinfo {author} {\bibfnamefont
			{L.}~\bibnamefont {Paulatto}}, \bibinfo {author} {\bibfnamefont
			{C.}~\bibnamefont {Sbraccia}}, \bibinfo {author} {\bibfnamefont
			{S.}~\bibnamefont {Scandolo}}, \bibinfo {author} {\bibfnamefont
			{G.}~\bibnamefont {Sclauzero}}, \bibinfo {author} {\bibfnamefont {A.~P.}\
			\bibnamefont {Seitsonen}}, \bibinfo {author} {\bibfnamefont {A.}~\bibnamefont
			{Smogunov}}, \bibinfo {author} {\bibfnamefont {P.}~\bibnamefont {Umari}},\
		and\ \bibinfo {author} {\bibfnamefont {R.~M.}\ \bibnamefont {Wentzcovitch}},\
	}\bibfield  {title} {\bibinfo {title} {{QUANTUM ESPRESSO: A Modular and
				Open-Source Software Project for Quantum Simulations of Materials}},\
	}\href@noop {} {\bibfield  {journal} {\bibinfo  {journal} {J. Phys.: Condens.
				Matterr}\ }\textbf {\bibinfo {volume} {21}},\ \bibinfo {pages} {395502}
		(\bibinfo {year} {2009})}\BibitemShut {NoStop}%
	\bibitem [{\citenamefont {Giannozzi}\ \emph {et~al.}(2017)\citenamefont
		{Giannozzi}, \citenamefont {Andreussi}, \citenamefont {Brumme}, \citenamefont
		{Bunau}, \citenamefont {Buongiorno~Nardelli}, \citenamefont {Calandra},
		\citenamefont {Car}, \citenamefont {Cavazzoni}, \citenamefont {Ceresoli},
		\citenamefont {Cococcioni}, \citenamefont {Colonna}, \citenamefont
		{Carnimeo}, \citenamefont {Dal~Corso}, \citenamefont {{de Gironcoli}},
		\citenamefont {Delugas}, \citenamefont {DiStasio}, \citenamefont {Ferretti},
		\citenamefont {Floris}, \citenamefont {Fratesi}, \citenamefont {Fugallo},
		\citenamefont {Gebauer}, \citenamefont {Gerstmann}, \citenamefont {Giustino},
		\citenamefont {Gorni}, \citenamefont {Jia}, \citenamefont {Kawamura},
		\citenamefont {Ko}, \citenamefont {Kokalj}, \citenamefont {K{\"u}{\c
				c}{\"u}kbenli}, \citenamefont {Lazzeri}, \citenamefont {Marsili},
		\citenamefont {Marzari}, \citenamefont {Mauri}, \citenamefont {Nguyen},
		\citenamefont {Nguyen}, \citenamefont {{Otero-de-la-Roza}}, \citenamefont
		{Paulatto}, \citenamefont {Ponc{\'e}}, \citenamefont {Rocca}, \citenamefont
		{Sabatini}, \citenamefont {Santra}, \citenamefont {Schlipf}, \citenamefont
		{Seitsonen}, \citenamefont {Smogunov}, \citenamefont {Timrov}, \citenamefont
		{Thonhauser}, \citenamefont {Umari}, \citenamefont {Vast}, \citenamefont
		{Wu},\ and\ \citenamefont {Baroni}}]{cite18}%
	\BibitemOpen
	\bibfield  {author} {\bibinfo {author} {\bibfnamefont {P.}~\bibnamefont
			{Giannozzi}}, \bibinfo {author} {\bibfnamefont {O.}~\bibnamefont
			{Andreussi}}, \bibinfo {author} {\bibfnamefont {T.}~\bibnamefont {Brumme}},
		\bibinfo {author} {\bibfnamefont {O.}~\bibnamefont {Bunau}}, \bibinfo
		{author} {\bibfnamefont {M.}~\bibnamefont {Buongiorno~Nardelli}}, \bibinfo
		{author} {\bibfnamefont {M.}~\bibnamefont {Calandra}}, \bibinfo {author}
		{\bibfnamefont {R.}~\bibnamefont {Car}}, \bibinfo {author} {\bibfnamefont
			{C.}~\bibnamefont {Cavazzoni}}, \bibinfo {author} {\bibfnamefont
			{D.}~\bibnamefont {Ceresoli}}, \bibinfo {author} {\bibfnamefont
			{M.}~\bibnamefont {Cococcioni}}, \bibinfo {author} {\bibfnamefont
			{N.}~\bibnamefont {Colonna}}, \bibinfo {author} {\bibfnamefont
			{I.}~\bibnamefont {Carnimeo}}, \bibinfo {author} {\bibfnamefont
			{A.}~\bibnamefont {Dal~Corso}}, \bibinfo {author} {\bibfnamefont
			{S.}~\bibnamefont {{de Gironcoli}}}, \bibinfo {author} {\bibfnamefont
			{P.}~\bibnamefont {Delugas}}, \bibinfo {author} {\bibfnamefont {R.~A.}\
			\bibnamefont {DiStasio}}, \bibinfo {author} {\bibfnamefont {A.}~\bibnamefont
			{Ferretti}}, \bibinfo {author} {\bibfnamefont {A.}~\bibnamefont {Floris}},
		\bibinfo {author} {\bibfnamefont {G.}~\bibnamefont {Fratesi}}, \bibinfo
		{author} {\bibfnamefont {G.}~\bibnamefont {Fugallo}}, \bibinfo {author}
		{\bibfnamefont {R.}~\bibnamefont {Gebauer}}, \bibinfo {author} {\bibfnamefont
			{U.}~\bibnamefont {Gerstmann}}, \bibinfo {author} {\bibfnamefont
			{F.}~\bibnamefont {Giustino}}, \bibinfo {author} {\bibfnamefont
			{T.}~\bibnamefont {Gorni}}, \bibinfo {author} {\bibfnamefont
			{J.}~\bibnamefont {Jia}}, \bibinfo {author} {\bibfnamefont {M.}~\bibnamefont
			{Kawamura}}, \bibinfo {author} {\bibfnamefont {H.-Y.}\ \bibnamefont {Ko}},
		\bibinfo {author} {\bibfnamefont {A.}~\bibnamefont {Kokalj}}, \bibinfo
		{author} {\bibfnamefont {E.}~\bibnamefont {K{\"u}{\c c}{\"u}kbenli}},
		\bibinfo {author} {\bibfnamefont {M.}~\bibnamefont {Lazzeri}}, \bibinfo
		{author} {\bibfnamefont {M.}~\bibnamefont {Marsili}}, \bibinfo {author}
		{\bibfnamefont {N.}~\bibnamefont {Marzari}}, \bibinfo {author} {\bibfnamefont
			{F.}~\bibnamefont {Mauri}}, \bibinfo {author} {\bibfnamefont {N.~L.}\
			\bibnamefont {Nguyen}}, \bibinfo {author} {\bibfnamefont {H.-V.}\
			\bibnamefont {Nguyen}}, \bibinfo {author} {\bibfnamefont {A.}~\bibnamefont
			{{Otero-de-la-Roza}}}, \bibinfo {author} {\bibfnamefont {L.}~\bibnamefont
			{Paulatto}}, \bibinfo {author} {\bibfnamefont {S.}~\bibnamefont {Ponc{\'e}}},
		\bibinfo {author} {\bibfnamefont {D.}~\bibnamefont {Rocca}}, \bibinfo
		{author} {\bibfnamefont {R.}~\bibnamefont {Sabatini}}, \bibinfo {author}
		{\bibfnamefont {B.}~\bibnamefont {Santra}}, \bibinfo {author} {\bibfnamefont
			{M.}~\bibnamefont {Schlipf}}, \bibinfo {author} {\bibfnamefont {A.~P.}\
			\bibnamefont {Seitsonen}}, \bibinfo {author} {\bibfnamefont {A.}~\bibnamefont
			{Smogunov}}, \bibinfo {author} {\bibfnamefont {I.}~\bibnamefont {Timrov}},
		\bibinfo {author} {\bibfnamefont {T.}~\bibnamefont {Thonhauser}}, \bibinfo
		{author} {\bibfnamefont {P.}~\bibnamefont {Umari}}, \bibinfo {author}
		{\bibfnamefont {N.}~\bibnamefont {Vast}}, \bibinfo {author} {\bibfnamefont
			{X.}~\bibnamefont {Wu}},\ and\ \bibinfo {author} {\bibfnamefont
			{S.}~\bibnamefont {Baroni}},\ }\bibfield  {title} {\bibinfo {title}
		{{Advanced Capabilities for Materials Modelling with Quantum ESPRESSO}},\
	}\href@noop {} {\bibfield  {journal} {\bibinfo  {journal} {J. Phys.: Condens.
				Matter}\ }\textbf {\bibinfo {volume} {29}},\ \bibinfo {pages} {465901}
		(\bibinfo {year} {2017})}\BibitemShut {NoStop}%
	\bibitem [{\citenamefont {Dal~Corso}(2014)}]{cite19}%
	\BibitemOpen
	\bibfield  {author} {\bibinfo {author} {\bibfnamefont {A.}~\bibnamefont
			{Dal~Corso}},\ }\bibfield  {title} {\bibinfo {title} {{Pseudopotentials
				Periodic Table: From H to Pu}},\ }\href@noop {} {\bibfield  {journal}
		{\bibinfo  {journal} {Comput. Mater. Sci.}\ }\textbf {\bibinfo {volume}
			{95}},\ \bibinfo {pages} {337} (\bibinfo {year} {2014})}\BibitemShut
	{NoStop}%
	\bibitem [{\citenamefont {Wannier}(1937)}]{cite20}%
	\BibitemOpen
	\bibfield  {author} {\bibinfo {author} {\bibfnamefont {G.~H.}\ \bibnamefont
			{Wannier}},\ }\bibfield  {title} {\bibinfo {title} {{The Structure of
				Electronic Excitation Levels in Insulating Crystals}},\ }\href@noop {}
	{\bibfield  {journal} {\bibinfo  {journal} {Phys. Rev.}\ }\textbf {\bibinfo
			{volume} {52}},\ \bibinfo {pages} {191} (\bibinfo {year} {1937})}\BibitemShut
	{NoStop}%
	\bibitem [{\citenamefont {Marzari}\ \emph {et~al.}(2012)\citenamefont
		{Marzari}, \citenamefont {Mostofi}, \citenamefont {Yates}, \citenamefont
		{Souza},\ and\ \citenamefont {Vanderbilt}}]{cite21}%
	\BibitemOpen
	\bibfield  {author} {\bibinfo {author} {\bibfnamefont {N.}~\bibnamefont
			{Marzari}}, \bibinfo {author} {\bibfnamefont {A.~A.}\ \bibnamefont
			{Mostofi}}, \bibinfo {author} {\bibfnamefont {J.~R.}\ \bibnamefont {Yates}},
		\bibinfo {author} {\bibfnamefont {I.}~\bibnamefont {Souza}},\ and\ \bibinfo
		{author} {\bibfnamefont {D.}~\bibnamefont {Vanderbilt}},\ }\bibfield  {title}
	{\bibinfo {title} {{Maximally Localized Wannier Functions: Theory and
				Applications}},\ }\href@noop {} {\bibfield  {journal} {\bibinfo  {journal}
			{Rev. Mod. Phys.}\ }\textbf {\bibinfo {volume} {84}},\ \bibinfo {pages}
		{1419} (\bibinfo {year} {2012})}\BibitemShut {NoStop}%
	\bibitem [{\citenamefont {Marzari}\ and\ \citenamefont
		{Vanderbilt}(1997)}]{cite22}%
	\BibitemOpen
	\bibfield  {author} {\bibinfo {author} {\bibfnamefont {N.}~\bibnamefont
			{Marzari}}\ and\ \bibinfo {author} {\bibfnamefont {D.}~\bibnamefont
			{Vanderbilt}},\ }\bibfield  {title} {\bibinfo {title} {{Maximally Localized
				Generalized Wannier Functions for Composite Energy Bands}},\ }\href@noop {}
	{\bibfield  {journal} {\bibinfo  {journal} {Phys. Rev. B}\ }\textbf {\bibinfo
			{volume} {56}},\ \bibinfo {pages} {12847} (\bibinfo {year}
		{1997})}\BibitemShut {NoStop}%
	\bibitem [{\citenamefont {Mostofi}\ \emph {et~al.}(2008)\citenamefont
		{Mostofi}, \citenamefont {Yates}, \citenamefont {Lee}, \citenamefont {Souza},
		\citenamefont {Vanderbilt},\ and\ \citenamefont {Marzari}}]{cite23}%
	\BibitemOpen
	\bibfield  {author} {\bibinfo {author} {\bibfnamefont {A.~A.}\ \bibnamefont
			{Mostofi}}, \bibinfo {author} {\bibfnamefont {J.~R.}\ \bibnamefont {Yates}},
		\bibinfo {author} {\bibfnamefont {Y.-S.}\ \bibnamefont {Lee}}, \bibinfo
		{author} {\bibfnamefont {I.}~\bibnamefont {Souza}}, \bibinfo {author}
		{\bibfnamefont {D.}~\bibnamefont {Vanderbilt}},\ and\ \bibinfo {author}
		{\bibfnamefont {N.}~\bibnamefont {Marzari}},\ }\bibfield  {title} {\bibinfo
		{title} {{Wannier90: A Tool for Obtaining Maximally-Localised Wannier
				Functions}},\ }\href@noop {} {\bibfield  {journal} {\bibinfo  {journal}
			{Comput. Phys. Commun.}\ }\textbf {\bibinfo {volume} {178}},\ \bibinfo
		{pages} {685} (\bibinfo {year} {2008})}\BibitemShut {NoStop}%
	\bibitem [{\citenamefont {Pizzi}\ \emph {et~al.}(2020)\citenamefont {Pizzi},
		\citenamefont {Vitale}, \citenamefont {Arita}, \citenamefont {Bl{\"u}gel},
		\citenamefont {Freimuth}, \citenamefont {G{\'e}ranton}, \citenamefont
		{Gibertini}, \citenamefont {Gresch}, \citenamefont {Johnson}, \citenamefont
		{Koretsune}, \citenamefont {{Iba{\~n}ez-Azpiroz}}, \citenamefont {Lee},
		\citenamefont {Lihm}, \citenamefont {Marchand}, \citenamefont {Marrazzo},
		\citenamefont {Mokrousov}, \citenamefont {Mustafa}, \citenamefont {Nohara},
		\citenamefont {Nomura}, \citenamefont {Paulatto}, \citenamefont {Ponc{\'e}},
		\citenamefont {Ponweiser}, \citenamefont {Qiao}, \citenamefont {Th{\"o}le},
		\citenamefont {Tsirkin}, \citenamefont {Wierzbowska}, \citenamefont
		{Marzari}, \citenamefont {Vanderbilt}, \citenamefont {Souza}, \citenamefont
		{Mostofi},\ and\ \citenamefont {Yates}}]{cite24}%
	\BibitemOpen
	\bibfield  {author} {\bibinfo {author} {\bibfnamefont {G.}~\bibnamefont
			{Pizzi}}, \bibinfo {author} {\bibfnamefont {V.}~\bibnamefont {Vitale}},
		\bibinfo {author} {\bibfnamefont {R.}~\bibnamefont {Arita}}, \bibinfo
		{author} {\bibfnamefont {S.}~\bibnamefont {Bl{\"u}gel}}, \bibinfo {author}
		{\bibfnamefont {F.}~\bibnamefont {Freimuth}}, \bibinfo {author}
		{\bibfnamefont {G.}~\bibnamefont {G{\'e}ranton}}, \bibinfo {author}
		{\bibfnamefont {M.}~\bibnamefont {Gibertini}}, \bibinfo {author}
		{\bibfnamefont {D.}~\bibnamefont {Gresch}}, \bibinfo {author} {\bibfnamefont
			{C.}~\bibnamefont {Johnson}}, \bibinfo {author} {\bibfnamefont
			{T.}~\bibnamefont {Koretsune}}, \bibinfo {author} {\bibfnamefont
			{J.}~\bibnamefont {{Iba{\~n}ez-Azpiroz}}}, \bibinfo {author} {\bibfnamefont
			{H.}~\bibnamefont {Lee}}, \bibinfo {author} {\bibfnamefont {J.-M.}\
			\bibnamefont {Lihm}}, \bibinfo {author} {\bibfnamefont {D.}~\bibnamefont
			{Marchand}}, \bibinfo {author} {\bibfnamefont {A.}~\bibnamefont {Marrazzo}},
		\bibinfo {author} {\bibfnamefont {Y.}~\bibnamefont {Mokrousov}}, \bibinfo
		{author} {\bibfnamefont {J.~I.}\ \bibnamefont {Mustafa}}, \bibinfo {author}
		{\bibfnamefont {Y.}~\bibnamefont {Nohara}}, \bibinfo {author} {\bibfnamefont
			{Y.}~\bibnamefont {Nomura}}, \bibinfo {author} {\bibfnamefont
			{L.}~\bibnamefont {Paulatto}}, \bibinfo {author} {\bibfnamefont
			{S.}~\bibnamefont {Ponc{\'e}}}, \bibinfo {author} {\bibfnamefont
			{T.}~\bibnamefont {Ponweiser}}, \bibinfo {author} {\bibfnamefont
			{J.}~\bibnamefont {Qiao}}, \bibinfo {author} {\bibfnamefont {F.}~\bibnamefont
			{Th{\"o}le}}, \bibinfo {author} {\bibfnamefont {S.~S.}\ \bibnamefont
			{Tsirkin}}, \bibinfo {author} {\bibfnamefont {M.}~\bibnamefont
			{Wierzbowska}}, \bibinfo {author} {\bibfnamefont {N.}~\bibnamefont
			{Marzari}}, \bibinfo {author} {\bibfnamefont {D.}~\bibnamefont {Vanderbilt}},
		\bibinfo {author} {\bibfnamefont {I.}~\bibnamefont {Souza}}, \bibinfo
		{author} {\bibfnamefont {A.~A.}\ \bibnamefont {Mostofi}},\ and\ \bibinfo
		{author} {\bibfnamefont {J.~R.}\ \bibnamefont {Yates}},\ }\bibfield  {title}
	{\bibinfo {title} {{Wannier90 as a Community Code: New Features and
				Applications}},\ }\href@noop {} {\bibfield  {journal} {\bibinfo  {journal}
			{J. Phys.: Condens. Matterr}\ }\textbf {\bibinfo {volume} {32}},\ \bibinfo
		{pages} {165902} (\bibinfo {year} {2020})}\BibitemShut {NoStop}%
	\bibitem [{\citenamefont {Ziman}(1972)}]{cite25}%
	\BibitemOpen
	\bibfield  {author} {\bibinfo {author} {\bibfnamefont {J.~M.}\ \bibnamefont
			{Ziman}},\ }\href {https://doi.org/10.1017/CBO9781139644075} {\emph {\bibinfo
			{title} {{Principles of the Theory of Solids}}}},\ \bibinfo {edition} {2nd}\
	ed.\ (\bibinfo  {publisher} {Cambridge University Press},\ \bibinfo {year}
	{1972})\BibitemShut {NoStop}%
	\bibitem [{\citenamefont {Grosso}\ and\ \citenamefont
		{Parravicini}(2014)}]{cite26}%
	\BibitemOpen
	\bibfield  {author} {\bibinfo {author} {\bibfnamefont {G.}~\bibnamefont
			{Grosso}}\ and\ \bibinfo {author} {\bibfnamefont {G.~P.}\ \bibnamefont
			{Parravicini}},\ }\href@noop {} {\emph {\bibinfo {title} {{Solid State
					Physics}}}},\ \bibinfo {edition} {2nd}\ ed.\ (\bibinfo  {publisher} {Academic
		Press},\ \bibinfo {address} {Amsterdam},\ \bibinfo {year} {2014})\BibitemShut
	{NoStop}%
	\bibitem [{\citenamefont {Mahan}(2006)}]{cite27}%
	\BibitemOpen
	\bibfield  {author} {\bibinfo {author} {\bibfnamefont {G.~D.}\ \bibnamefont
			{Mahan}},\ }\bibfield  {title} {\bibinfo {title} {{Transport Properties}},\
	}in\ \href@noop {} {\emph {\bibinfo {booktitle} {International Tables for
				Crystallography}}},\ Vol.~\bibinfo {volume} {D}\ (\bibinfo {year} {2006})\
	Chap.\ \bibinfo {chapter} {1.8}, pp.\ \bibinfo {pages} {220--227}\BibitemShut
	{NoStop}%
	\bibitem [{\citenamefont {Pizzi}\ \emph {et~al.}(2013)\citenamefont {Pizzi},
		\citenamefont {Volja}, \citenamefont {Kozinsky}, \citenamefont {Fornari},\
		and\ \citenamefont {Marzari}}]{cite28}%
	\BibitemOpen
	\bibfield  {author} {\bibinfo {author} {\bibfnamefont {G.}~\bibnamefont
			{Pizzi}}, \bibinfo {author} {\bibfnamefont {D.}~\bibnamefont {Volja}},
		\bibinfo {author} {\bibfnamefont {B.}~\bibnamefont {Kozinsky}}, \bibinfo
		{author} {\bibfnamefont {M.}~\bibnamefont {Fornari}},\ and\ \bibinfo {author}
		{\bibfnamefont {N.}~\bibnamefont {Marzari}},\ }\bibfield  {title} {\bibinfo
		{title} {{BoltzWann: A code for the evaluation of thermoelectric and
				electronic transport properties with a maximally-localized Wannier functions
				basis}},\ }\href@noop {} {\bibfield  {journal} {\bibinfo  {journal} {Comput.
				Phys. Commun.}\ }\textbf {\bibinfo {volume} {185}},\ \bibinfo {pages} {422}
		(\bibinfo {year} {2013})}\BibitemShut {NoStop}%
	\bibitem [{\citenamefont {Monchesky}\ \emph {et~al.}(2005)\citenamefont
		{Monchesky}, \citenamefont {Enders}, \citenamefont {Urban}, \citenamefont
		{Myrtle}, \citenamefont {Heinrich}, \citenamefont {Zhang}, \citenamefont
		{Butler},\ and\ \citenamefont {Kirschner}}]{cite29}%
	\BibitemOpen
	\bibfield  {author} {\bibinfo {author} {\bibfnamefont {T.~L.}\ \bibnamefont
			{Monchesky}}, \bibinfo {author} {\bibfnamefont {A.}~\bibnamefont {Enders}},
		\bibinfo {author} {\bibfnamefont {R.}~\bibnamefont {Urban}}, \bibinfo
		{author} {\bibfnamefont {K.}~\bibnamefont {Myrtle}}, \bibinfo {author}
		{\bibfnamefont {B.}~\bibnamefont {Heinrich}}, \bibinfo {author}
		{\bibfnamefont {X.-G.}\ \bibnamefont {Zhang}}, \bibinfo {author}
		{\bibfnamefont {W.~H.}\ \bibnamefont {Butler}},\ and\ \bibinfo {author}
		{\bibfnamefont {J.}~\bibnamefont {Kirschner}},\ }\bibfield  {title} {\bibinfo
		{title} {{Spin-dependent transport in Fe and Fe/Au multilayers}},\ }\href
	{https://doi.org/10.1103/PhysRevB.71.214440} {\bibfield  {journal} {\bibinfo
			{journal} {Phys. Rev. B}\ }\textbf {\bibinfo {volume} {71}},\ \bibinfo
		{pages} {214440} (\bibinfo {year} {2005})}\BibitemShut {NoStop}%
	\bibitem [{\citenamefont {Nagaosa}\ \emph {et~al.}(2010)\citenamefont
		{Nagaosa}, \citenamefont {Sinova}, \citenamefont {Onoda}, \citenamefont
		{MacDonald},\ and\ \citenamefont {Ong}}]{cite30}%
	\BibitemOpen
	\bibfield  {author} {\bibinfo {author} {\bibfnamefont {N.}~\bibnamefont
			{Nagaosa}}, \bibinfo {author} {\bibfnamefont {J.}~\bibnamefont {Sinova}},
		\bibinfo {author} {\bibfnamefont {S.}~\bibnamefont {Onoda}}, \bibinfo
		{author} {\bibfnamefont {A.~H.}\ \bibnamefont {MacDonald}},\ and\ \bibinfo
		{author} {\bibfnamefont {N.~P.}\ \bibnamefont {Ong}},\ }\bibfield  {title}
	{\bibinfo {title} {{Anomalous Hall effect}},\ }\href
	{https://doi.org/10.1103/RevModPhys.82.1539} {\bibfield  {journal} {\bibinfo
			{journal} {Rev. Mod. Phys.}\ }\textbf {\bibinfo {volume} {82}},\ \bibinfo
		{pages} {1539} (\bibinfo {year} {2010})}\BibitemShut {NoStop}%
	\bibitem [{\citenamefont {Li}\ \emph {et~al.}(2023)\citenamefont {Li},
		\citenamefont {Zhang}, \citenamefont {Guo}, \citenamefont {Min},\ and\
		\citenamefont {Wang}}]{cite31}%
	\BibitemOpen
	\bibfield  {author} {\bibinfo {author} {\bibfnamefont {P.}~\bibnamefont
			{Li}}, \bibinfo {author} {\bibfnamefont {J.-Z.}\ \bibnamefont {Zhang}},
		\bibinfo {author} {\bibfnamefont {Z.-X.}\ \bibnamefont {Guo}}, \bibinfo
		{author} {\bibfnamefont {T.}~\bibnamefont {Min}},\ and\ \bibinfo {author}
		{\bibfnamefont {X.}~\bibnamefont {Wang}},\ }\bibfield  {title} {\bibinfo
		{title} {{Intrinsic Anomalous Spin Hall Effect}},\ }\href@noop {} {\bibfield
		{journal} {\bibinfo  {journal} {Sci. China: Phys., Mech. Astron.}\ }\textbf
		{\bibinfo {volume} {66}},\ \bibinfo {pages} {227511} (\bibinfo {year}
		{2023})}\BibitemShut {NoStop}%
	\bibitem [{\citenamefont {Campbell}\ \emph
		{et~al.}(1970{\natexlab{b}})\citenamefont {Campbell}, \citenamefont {Fert},\
		and\ \citenamefont {Jaoul}}]{cite32}%
	\BibitemOpen
	\bibfield  {author} {\bibinfo {author} {\bibfnamefont {I.~A.}\ \bibnamefont
			{Campbell}}, \bibinfo {author} {\bibfnamefont {A.}~\bibnamefont {Fert}},\
		and\ \bibinfo {author} {\bibfnamefont {O.}~\bibnamefont {Jaoul}},\ }\bibfield
	{title} {\bibinfo {title} {{The spontaneous resistivity anisotropy in
				Ni-based alloys}},\ }\href {https://doi.org/10.1088/0022-3719/3/1S/310}
	{\bibfield  {journal} {\bibinfo  {journal} {J. Phys. C: Solid State Phys.}\
		}\textbf {\bibinfo {volume} {3}},\ \bibinfo {pages} {S95} (\bibinfo {year}
		{1970}{\natexlab{b}})}\BibitemShut {NoStop}%
	\bibitem [{\citenamefont {Kokado}\ \emph {et~al.}(2012)\citenamefont {Kokado},
		\citenamefont {Tsunoda}, \citenamefont {Harigaya},\ and\ \citenamefont
		{Sakuma}}]{cite33}%
	\BibitemOpen
	\bibfield  {author} {\bibinfo {author} {\bibfnamefont {S.}~\bibnamefont
			{Kokado}}, \bibinfo {author} {\bibfnamefont {M.}~\bibnamefont {Tsunoda}},
		\bibinfo {author} {\bibfnamefont {K.}~\bibnamefont {Harigaya}},\ and\
		\bibinfo {author} {\bibfnamefont {A.}~\bibnamefont {Sakuma}},\ }\bibfield
	{title} {\bibinfo {title} {{Anisotropic Magnetoresistance Effects in Fe, Co,
				Ni, Fe\textsubscript{4}N, and Half-Metallic Ferromagnet: A Systematic
				Analysis}},\ }\href {https://doi.org/10.1143/JPSJ.81.024705} {\bibfield
		{journal} {\bibinfo  {journal} {J. Phys. Soc. Jpn.}\ }\textbf {\bibinfo
			{volume} {81}},\ \bibinfo {pages} {024705} (\bibinfo {year}
		{2012})}\BibitemShut {NoStop}%
	\bibitem [{\citenamefont {Kokado}\ and\ \citenamefont
		{Tsunoda}(2013)}]{cite34}%
	\BibitemOpen
	\bibfield  {author} {\bibinfo {author} {\bibfnamefont {S.}~\bibnamefont
			{Kokado}}\ and\ \bibinfo {author} {\bibfnamefont {M.}~\bibnamefont
			{Tsunoda}},\ }\bibfield  {title} {\bibinfo {title} {{Anisotropic
				Magnetoresistance Effect: General Expression of AMR Ratio and Intuitive
				Explanation for Sign of AMR Ratio}},\ }in\ \href
	{https://doi.org/10.4028/www.scientific.net/AMR.750-752.978} {\emph {\bibinfo
			{booktitle} {Advanced Engineering Materials III}}},\ \bibinfo {series} {Adv.
		Mater. Res.}, Vol.\ \bibinfo {volume} {750}\ (\bibinfo  {publisher} {Trans
		Tech Publications Ltd},\ \bibinfo {year} {2013})\ pp.\ \bibinfo {pages}
	{978--982}\BibitemShut {NoStop}%
	\bibitem [{\citenamefont {Wang}\ \emph {et~al.}(2023)\citenamefont {Wang},
		\citenamefont {Wang},\ and\ \citenamefont {Wang}}]{XRW}%
	\BibitemOpen
	\bibfield  {author} {\bibinfo {author} {\bibfnamefont {X.~R.}\ \bibnamefont
			{Wang}}, \bibinfo {author} {\bibfnamefont {C.}~\bibnamefont {Wang}},\ and\
		\bibinfo {author} {\bibfnamefont {X.~S.}\ \bibnamefont {Wang}},\ }\bibfield
	{title} {\bibinfo {title} {{A theory of unusual anisotropic magnetoresistance
				in bilayer heterostructures}},\ }\href@noop {} {\bibfield  {journal}
		{\bibinfo  {journal} {Sci. Rep.}\ }\textbf {\bibinfo {volume} {13}},\
		\bibinfo {pages} {309} (\bibinfo {year} {2023})}\BibitemShut {NoStop}%
\end{thebibliography}
%

\end{document}